\documentclass[a4paper,11pt]{article}

\usepackage [cp 1251]{inputenc}
\usepackage [russian] {babel}

\usepackage{amssymb,amsbsy}
\usepackage{amsmath,amscd}
\usepackage{amssymb}
\usepackage{amsmath}
\usepackage{amsfonts}
\usepackage{color}
\usepackage{mathrsfs, graphicx}
\DeclareMathAccent{\wtilde}{\mathord}{largesymbols}{"65}
\DeclareMathAccent{\what}{\mathord}{largesymbols}{"62}

\newcommand\cD{{\mathcal D}}

\newcommand\bbbZ{{\mathbb Z}}

\newcommand\bbbK{{\mathbb K}}

\newcommand{\gA}{{\mathfrak{A}}}

\newcommand{\gJ}{{\mathfrak{J}}}

\newcommand{\ccdot}{{ }}

\newcommand\ccS{{S}}

\newtheorem{prop}{Proposition}

\title{Quantisation ideals of nonabelian integrable systems}
\author{A.V.Mikhailov}
\date{}

\begin{document}

\maketitle
We consider dynamical systems on the space of functions taking values in a free associative algebra \cite{EGR98, miksok_CMP}. The system is said to be integrable if it possesses an infinite dimensional Lie algebra of commuting symmetries. In this paper we propose a new approach to the problem of quantisation of dynamical systems, introduce the concept of quantisation ideals and provide meaningful examples. 

In order to illustrate the new approach we have chosen  the following nonabelian integrable systems: the Volterra chain (i)  and the Bogoyavlensky  $N$--chains (ii) \cite{bog}
\begin{equation}\label{bog}
({\rm i})\  u_t=u_{1}\ccdot u-u \ccdot u_{-1},\quad ({\rm ii})\ \frac{du}{dt}=\sum_{k=1}^N (u_k\ccdot u-u\ccdot u_{-k}),
\end{equation}
These are infinite systems of equations where we use standard notations  $u=u_0=u(n,t),\ u_k=u_k(n,t)=u(n+k,t),\ n,k\in\bbbZ$. In equations (\ref{bog}) functions $u_k$ are elements of a free associative algebra  $\gA=\bbbK \langle\dots u_{-1},u,u_1,\ldots\rangle$ over a zero characteristic field of constants $\bbbK$.

Traditional quantisation theories start with classical systems on functions taking values in commutative algebras. In this paper we propose a new approach departing from the systems  defined on free associative algebras. In this approach the quantisation problem is reduced to a description  of two-sided ideals which define the commutation relations in the quotient algebras and are invariant with respect to the dynamics of the system. We begin with consideration of two-sided ideals
 $\gJ_F \subset\gA$ generated by an infinite set of polynomials of the form
\begin{equation}\label{F}
\gJ_F=\langle\{F_{p,q}=u_qu_p-\omega_{p,q}u_p u_q\,|\, p,q\in\bbbZ,\ p>q,\ \omega_{p,q}\in\bbbK^*\} \rangle
\end{equation}
and find out such structure constants  $\omega_{p,q}$, that equations (\ref{bog}) 
could be restricted to the quotient algebra $\gA_{\gJ_F}=\gA\diagup{\gJ_F}$. In such a case we say that system  (\ref{bog}) admits $\gJ_F${\em --quantisation} and is defined on the  {\em quantised algebra} $\gA_{\gJ_F}$.

In general, by {\em quantised algebra} it is understood such a non-commutative quotient algebra $\gA_\gJ=\gA\diagup\gJ$ that has an additive basis of lexicographically  ordered monomials of the form   $u_{i_1}^{a_1}u_{i_2}^{a_2}\cdots u_{i_n}^{a_n}$ in finite or infinite number of variables, where   $i_1<i_2<\cdots<i_n,\ i_k\in\bbbZ$. The ideal 
$\gJ$ defines commutation relations in the quotient algebra and we call it the {\em quantisation ideal}. In the case of finitely generated algebras it is said that $\gA_{\gJ}$ has a Poincar\'e--Birkhoff-Witt basis.

Algebra $\gA$ has a natural automorphism generated by the shift operator $\ccS(u_k)=u_{k+1},\ k\in\bbbZ$. Derivations 
$D_f$ of algebra $\gA$, commuting with the automorphism $\ccS$, are called evolutionary. It is sufficient to define an evolutionary derivation on one algebra generator $D_f(u)=f\in \gA$. The evolutionary derivation $D_f(u)$ is in one-to-one correspondence with the system of differential-difference equations
\begin{equation}\label{dd}
 \frac{du_k}{dt_f}=\ccS^k f
 \,\qquad f=f(\ldots,u_{-1},u,u_1,\ldots)\in\gA,\ \ k\in\bbbZ.
\end{equation}
By a symmetry of equation (\ref{dd}) we understand an evolutionary derivation  $D_g$which commutes with  $D_f$, or, in other words, the evolutionary differential-difference system of equations  $u_{t_g}=g$, which is compatible with (\ref{dd}). Symmetries $D_{g_1},D_{g_2}$ are called commuting, if $[D_{g_1},D_{g_2}]=0$. 
A derivation  $D_f$ of the algebra $\gA$, which maps the ideal $\gJ\subset\gA$ into itself $D_f(\gJ)\subseteq\gJ$, induces canonically a derivation  $\cD_f$ on $\gA_\gJ$. If  $\ccS^\ell(\gJ)\subseteq \gJ$  for a certain integer number $\ell>0$, then $\ccS^\ell$ is automorphism  of $\gA_\gJ$ commuting with $\cD_f$. It enables us to restrict system  (\ref{dd}) to  $\gA_\gJ$.

Let us begin with non-abelian Volterra system (\ref{bog}) (i) and its symmetry
\begin{equation}
\label{voltf2}
\frac{du}{d \tau}=u \ccdot u_{-1}\ccdot u_{-2}+u \ccdot u_{-1}\ccdot u_{-1}+u \ccdot u \ccdot u_{-1}-u_{1}\ccdot u \ccdot u -u_{1}\ccdot u_{1}\ccdot u -u_{2}\ccdot u_{1}\ccdot u .
\end{equation}
\begin{prop} Equation {\rm (\ref{voltf2})} can be restricted to $\gA_{\gJ_F}$ only in the following cases:
\[
 {\rm (i)}\ \ \omega_{n+1,n}=\alpha,\ \omega_{n,m}=1,\quad {\rm (ii)}\ \ \omega_{n+1,n}=(-1)^n \alpha,\ \omega_{n,m}=-1;\quad  n-m\ge 2 .
\]
Nonabelian Volterra system can be restricted to the algebra $\gA_{\gJ_F}$ only in the case {\rm (i)}.
\end{prop}
There are reasons to state that all symmetries of the Volterra hierarchy admit quantisation {\rm (i)} for which $\ccS(\gJ_F)=\gJ_F$. If we restrict ourself with odd-degree equations from the hierarchy, then there exists an alternative quantisation  {\rm (ii)} for which $\ccS^2(\gJ_F)= \gJ_F$. 
\begin{prop}
Nonabelian $N$--chain  {\rm (\ref{bog}) (ii)} admits $\gJ_F$--quantisation only in the case 
\[
 \omega_{n+k,n} =\alpha,\ \  \mbox{where}  \ \ 1\le k\le N,\ \alpha\ne 0,\ \mbox{and}\ \quad \omega_{n,m}=1,\ \ \mbox{for} \ \   n-m> N.
\]
\end{prop}
\begin{prop} There exists a modification 
\begin{equation}\label{mb2}
u_t=u_{2} \ccdot u_{1} \ccdot u \ccdot u + u_{1} \ccdot u \ccdot u_{-1} \ccdot u 
 - u \ccdot u_{1} \ccdot u \ccdot u_{-1}-u \ccdot u \ccdot u_{-1} \ccdot u_{-2}
\end{equation}
of the nonabelian  $N=2$  Bogoyavlensky chain. System (\ref{mb2}) admits $\gJ_F$--quantisation only in the case
\[ \omega_{3n+1,3m} =\alpha,\quad \omega_{3n+2,3m}=\beta,\quad \omega_{3n+3,3m} =\alpha^{-1}\beta^{-1},\qquad   \ \alpha,\beta\ne 0,\ n\ge m,\ n,m\in\bbbZ.
\]
\end{prop}
The quantisation of the chain  (\ref{mb2}) is defined by two independent ``deformation'' parameters $\alpha, \beta$. For bigger values of $N$   the number of independent parameters may also increase.  In the case  $N=3$ a quantisation of the corresponding modified chain is defined by four independent parameters. 

Periodic closures of the chains $u_{k+M}=u_{k}$ with  period  $M$ result in nonabelian systems of ordinary differential equations for functions taking their values in a finitely generated unital free algebra $\gA^M =\bbbK\langle u_1,\ldots, u_M\rangle$. In these cases the system may admit inhomogeneous quantisation ideals of the form 
\[
 \gJ_G=\langle \{G_{p,q}=u_qu_p-\omega_{p,q}u_p u_q+\sum_{r=1}^M \sigma^r_{p,q}u_r +\eta_{p,q}\,|\, p,q\in\bbbZ,\ p>q,\ \omega_{p,q}\ne 0, \omega_{p,q},\sigma^r_{p,q},\eta_{p,q}\in\bbbK\}\rangle
\]
\begin{prop} Nonabelian periodical Volterra chain {\rm (\ref{bog})  (i)} with  period  $M$ admits $\gJ_G$--quantisation iff the following commutation relations 
 \[
  \begin{array}{llll}
  M=2:& uu_1=\alpha u_1 u+\beta u_1+\gamma u+\eta;& M=3:& u_nu_{n+1}=\alpha u_{n+1} u_n+\beta (u+u_1+u_2)+\eta,\ \ n\in \bbbZ_3;\\ &&& \\
  M=4:&  u_1u_{2}=\alpha u_{2} u_1+\beta u_{2}+\gamma u_{1}-\beta \gamma,& & u_1 u_3=u_3 u_1-\beta u_2+\beta u_4,\\
 &u_4u_{1}=\alpha u_{1} u_4+\beta u_{4}+\gamma u_{1}-\beta \gamma,& &
 u_2u_{3}=\alpha u_{3} u_2+\beta u_{2}+\gamma u_{3}-\beta \gamma,\\
&u_2 u_4=u_4 u_2-\gamma u_3+\gamma u_1,& &
u_3u_{4}=\alpha u_{4} u_3+\beta u_{4}+\gamma u_{3}-\beta \gamma;\\&&&\\
 M\ge 5:&   u_{n+1}u_n=\alpha u_n u_{n+1},&& u_n u_m=u_mu_n,\ \ |n-m|>1,\ n,m\in\bbbZ_M.
 \end{array}
\]
take place. The constants $\alpha,\beta,\gamma,\eta\in\bbbK, \ \alpha\ne 0$ are arbitrary.
\end{prop}

The author is grateful to V.M.Buchstaber and V.V.Sokolov for useful discussions, and the  EPSRC grant EP/P012655/1 for  partial support.

School of Mathematics, University of Leeds, UK\\
a.v.mikhailov@leeds.ac.uk
\end{document}